\documentclass[twocolumn,secnumarabic,amssymb,nobibnotes,aps,prd]{revtex4-1}

\usepackage{color}
\usepackage{threeparttable}
\usepackage{srcltx}

\usepackage{graphicx,amsmath,psfrag,bm}
\usepackage[usenames,dvipsnames]{xcolor}
\usepackage{subfig}
\usepackage{cancel}
\usepackage{epstopdf}
\usepackage{pstool}

\newcommand{\ve}[1]{\boldsymbol{#1}}
\newcommand{\te}[1]{\overline{\overline{#1}}}

\begin{document}

\title{Mathematical Synthesis and Analysis of Nonlinear Metasurfaces}
\author{Karim~Achouri, Yousef Vahabzadeh, and~Christophe~Caloz}

\email[E-mail: ]{karim.achouri@polymtl.ca, christophe.caloz@polymtl.ca}
\affiliation{Department of Electrical Engineering, Polytechnique Montr\'{e}al, Montr\'{e}al, Qu\'{e}bec, Canada}

\begin{abstract}
We propose a discussion on the synthesis and scattering analysis of nonlinear metasurfaces. For simplicity, we investigate the case of a second-order nonlinear isotropic metasurface possessing both electric and magnetic linear and nonlinear susceptibility components. We next find the synthesis expressions relating the susceptibilities to the specified fields, which leads to the definition of the nonlinear metasurface conditions for no reflection, themselves revealing the nonreciprocal nature of such structures. Finally, we provide the approximate expressions of the scattered fields based on perturbation theory and compare the corresponding results to finite-difference time-domain simulations.
\end{abstract}

\maketitle

\section{Introduction}

Over the past few years, metasurfaces, the two-dimensional counterparts of three-dimensional metamaterials, have proven to be particularly effective at controlling electromagnetic waves. However, most studies on metasurfaces have been restricted to purely linear structures and only few studies, as for instance~\cite{lee2014giant,wakatsuchi2013waveform,chen2010optical,yang2015nonlinear}, have investigated the synthesis and/or scattering from nonlinear metasurfaces, but without providing extensive discussion on the topic. Since nonlinearity may potentially bring about a wealth of new applications to the realm of metasurface-based effects, such as for instance nonreciprocity, second-harmonic generation and wave-mixing~\cite{boyd2003nonlinear}, we propose here a rigorous discussion on the synthesis and scattering analysis from second-order nonlinear metasurfaces.

In the following, we will use the generalized sheet transition conditions (GSTCs) to obtain the metasurface susceptibilities (linear and nonlinear components) in terms of specified incident, reflected and transmitted waves. Based on that, the conditions for no reflection for second-order nonlinear metasurfaces will be derived, which will be next used to analyze the scattering from such structures. The scattered field will be computed using perturbation analysis and the results will be compared with FDTD simulations.

\section{Synthesis of Second-Order Nonlinear Metasurfaces}

The GSTCs are boundary conditions that apply to zero-thickness discontinuities such as metasurfaces~\cite{Idemen1973,kuester2003av,achouri2014general}. These conditions relate the discontinuities of the electric and magnetic fields to the presence of excitable surface polarization densities. In the case of a metasurface lying in the $xy$-plane at $z=0$ and assuming only transverse polarizations, the GSTCs read~\footnote{We assume here the harmonic time dependence $e^{j\omega t}$.}
\begin{subequations}
\label{eq:BC}
\begin{align}
\hat{z}\times\Delta\ve{H}
&=j\omega\ve{P}_{\parallel},\label{eq:CurlH}\\
\Delta\ve{E}\times\hat{z}
&=j\omega\mu \ve{M}_{\parallel},\label{eq:CurlE}
\end{align}
\end{subequations}
where $\Delta$ indicates the difference of the fields between both sides of the metasurface.

Let us now investigate the case of a metasurface with nonzero second-order nonlinear electric and magnetic susceptibility tensors. The presence of non-negligible electric nonlinearities may be found, for instance, in optical nonlinear crystals~\cite{boyd2003nonlinear} while magnetic nonlinearities may be found in ferrofluids~\cite{wang2006nonlinear}. We restrict our attention to the case of monoanisotropic metasurfaces~\cite{achouri2014general}, whose electric and magnetic polarization densities are
\begin{subequations}\label{eq:NLPol}
\begin{align}
\ve{P} &= \epsilon_0\te{\chi}^{(1)}_\text{ee}\ve{E}_\text{av} + \epsilon_0\te{\chi}^{(2)}_\text{ee}\ve{E}_\text{av}^2,\\
\ve{M} &= \te{\chi}^{(1)}_\text{mm}\ve{H}_\text{av} + \te{\chi}^{(2)}_\text{mm}\ve{H}_\text{av}^2,
\end{align}
\end{subequations}
where $\te{\chi}^{(1)}$ and $\te{\chi}^{(2)}$ correspond to the first-order (linear) and second-order (nonlinear) susceptibility tensors, respectively, and the subscript ``av'' denotes the average of the field between both sides of the metasurface. For simplicity, we assume that the only nonzero components of these tensors are $\te{\chi}^{aa,(1)}$ and $\te{\chi}^{aaa,(2)}$, where $a = \{x,y\}$, which generallly corresponds to the case of a birefringent nonlinear metasurface. Since nonlinear media generate new frequencies~\cite{boyd2003nonlinear}, the frequency-domain GSTCs in~\eqref{eq:BC} are not appropriate to investigate the response of nonlinear metasurfaces since they relate electromagnetic fields with the same frequency. To overcome this issue, we express the GSTCs in the time-domain instead of the frequency-domain. To further simplify the discussion, we consider, without loss of generality, only the case of $x$-polarized waves which, upon insertion of~\eqref{eq:NLPol} into the time-domain version of~\eqref{eq:BC}, reduces the time-domain GSTCs to
\begin{subequations}\label{eq:TDgstc}
\begin{align}
-\Delta H &= \epsilon_0 \chi^{(1)}_\text{ee} \frac{\partial}{\partial t} E_\text{av} + \epsilon_0 \chi^{(2)}_\text{ee} \frac{\partial}{\partial t} E^2_\text{av},\\
-\Delta E &= \mu_0 \chi^{(1)}_\text{mm} \frac{\partial}{\partial t} H_\text{av} + \mu_0 \chi^{(2)}_\text{mm} \frac{\partial}{\partial t} H^2_\text{av}\label{eq:TDgstc2},
\end{align}
\end{subequations}
where $E$ and $H$ are, respectively, the $x$-component of the electric field and the $y$-component of the magnetic field, and where the susceptibility components are those corresponding to $x$-polarized excitation.

To synthesize a nonlinear metasurface, one needs to solve~\eqref{eq:TDgstc} so as to express the susceptibilities as functions of the electromagnetic fields on both sides of the metasurface. As it stands in~\eqref{eq:TDgstc}, the system has two equations for four unknowns, and is hence under-determined. If we consider \emph{two} arbitrary transformations instead of just one~\cite{achouri2014general}, then the system becomes a full-rank one, reading
\begin{equation}
\begin{split}
&\begin{bmatrix}
-\Delta H_1 \\
-\Delta H_2 \\
-\Delta E_1 \\
-\Delta E_2
\end{bmatrix}=\\
\frac{\partial}{\partial t}
&\begin{bmatrix}
\epsilon_0  E_\text{av,1}   && \epsilon_0 E^2_\text{av,1}  && 0 && 0\\
\epsilon_0  E_\text{av,2}   && \epsilon_0 E^2_\text{av,2}  && 0 && 0\\
0 && 0 && \mu_0  H_\text{av,1}   && \mu_0 H^2_\text{av,1} \\
0 && 0 && \mu_0  H_\text{av,2}   && \mu_0 H^2_\text{av,2}
\end{bmatrix}
\begin{bmatrix}
\chi^{(1)}_\text{ee}\\
\chi^{(2)}_\text{ee}\\
\chi^{(1)}_\text{mm}\\
\chi^{(2)}_\text{mm}
\end{bmatrix},
\end{split}
\end{equation}
where the subscripts 1 and 2 refer to the fields of two arbitrary transformations. This matrix system is easily solved and yields the following expressions for the susceptibilities
\begin{subequations}\label{eq:NLX}
\begin{align}
\chi^{(1)}_\text{ee} &= -\frac{\Delta H_2 \frac{\partial}{\partial t} E^2_\text{av,1} - \Delta H_1 \frac{\partial}{\partial t} E^2_\text{av,2} }{\epsilon_0 (\frac{\partial}{\partial t} E^2_\text{av,1} \frac{\partial}{\partial t} E_\text{av,2} -\frac{\partial}{\partial t} E_\text{av,1} \frac{\partial}{\partial t} E^2_\text{av,2} )},\\
\chi^{(1)}_\text{mm} &= -\frac{\Delta E_2 \frac{\partial}{\partial t} H^2_\text{av,1} - \Delta E_1 \frac{\partial}{\partial t} H^2_\text{av,2} }{\mu_0 (\frac{\partial}{\partial t} H^2_\text{av,1} \frac{\partial}{\partial t} H_\text{av,2} -\frac{\partial}{\partial t} H_\text{av,1} \frac{\partial}{\partial t} H^2_\text{av,2} )},\\
\chi^{(2)}_\text{ee} &= \frac{\Delta H_2 \frac{\partial}{\partial t} E_\text{av,1} - \Delta H_1 \frac{\partial}{\partial t} E_\text{av,2} }{\epsilon_0 (\frac{\partial}{\partial t} E^2_\text{av,1} \frac{\partial}{\partial t} E_\text{av,2} -\frac{\partial}{\partial t} E_\text{av,1} \frac{\partial}{\partial t} E^2_\text{av,2} )},\\
\chi^{(2)}_\text{mm} &= \frac{\Delta E_2 \frac{\partial}{\partial t} H_\text{av,1} - \Delta E_1 \frac{\partial}{\partial t} H_\text{av,2} }{\mu_0 (\frac{\partial}{\partial t} H^2_\text{av,1} \frac{\partial}{\partial t} H_\text{av,2} -\frac{\partial}{\partial t} H_\text{av,1} \frac{\partial}{\partial t} H^2_\text{av,2} )}.
\end{align}
\end{subequations}

At this stage, one may think that substituting the specified arbitrary incident, reflected and transmitted fields into~\eqref{eq:NLX} would lead to well-defined susceptibilities. However, the specification of even ``simple'' transformations as, for instance, $E_\text{i} = E_0 e^{j(\omega_\text{i}t-k_\text{i}z)}$, $E_\text{r} = R e^{j(\omega_\text{r}t-k_\text{r}z)}$ and $E_\text{t} = T e^{j(\omega_\text{t}t-k_\text{t}z)}$, leads to susceptibilities that are \emph{time-varying}, irrespectively of the values of $\omega_\text{i}$, $\omega_\text{r}$ and $\omega_\text{t}$. The fact that the susceptibilities are, in this case, time-varying is inconsistent with the implicit assumption in Eqs.~\eqref{eq:TDgstc} that they are not~\footnote{If the susceptilities were time-dependent, insertion of~\eqref{eq:NLPol} into the time-domain version of~\eqref{eq:BC} would lead, from the derivation chain rule, to additional terms of the form $\left(\partial\chi_{uu}^{(k)}/\partial t\right)E_\text{av},H_\text{av}$, where $u=$e,m and $k=1,2$.}. At this point, one may therefore wonder whether the problem has not been inadequately paused? However, fortunately, we shall see in the next section that this is not the case: the problem is adequately posed but the specified fields must satisfy a specific constraint, while being otherwise still synthesizable. This constraint will be established by considering specified fields satisfying the postulate that the susceptibilities in~\eqref{eq:NLX} be not functions of time. For this purpose, we will look at the problem from a different perspective. Instead of trying to synthesize the metasurface, we shall heuristically analyze its scattering with (arbitrary) known susceptibilities in order to understand what kind of reflected and transmitted fields are produced by such nonlinear metasurfaces, and hence deduce a proper way to perform the synthesis.

\section{Scattering From Second-Order Nonlinear Metasurfaces}

\begin{figure}[ht]
\centering
\subfloat[]{\label{fig:NL1}
\includegraphics[width=0.45\columnwidth]{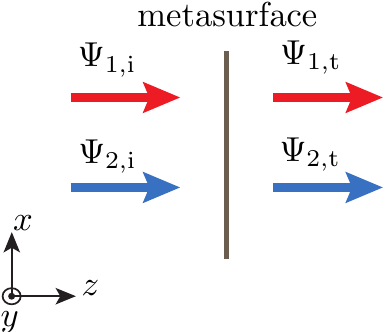}
}
\subfloat[]{\label{fig:NL2}
\includegraphics[width=0.45\columnwidth]{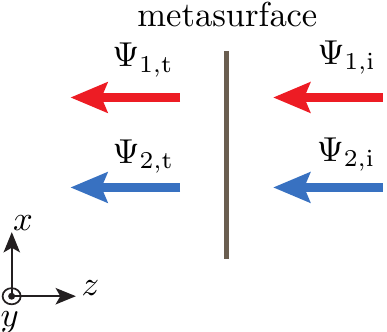}
}
\caption{Two different approaches to synthesize a nonlinear reflectionless metasurface. (a)~Metasurface reflectionlessly transmitting two waves ($\Psi_1$ and $\Psi_2$) in the $+z$-direction. (b)~Metasurface reflectionlessly transmitting two waves ($\Psi_1$ and $\Psi_2$) in the $-z$-direction.}
\label{fig:NL}
\end{figure}

The fields scattered from a nonlinear metasurface may be obtained by solving~\eqref{eq:TDgstc}. However, Eqs.~\eqref{eq:TDgstc} form a set of nonlinear inhomogeneous first-order coupled differential equations, that is not trivial to solve analytically. The problem may be simplified by assuming that the metasurface is reflectionless, which reduces~\eqref{eq:TDgstc} to a single equation.
The conditions for no reflection in a nonlinear metasurface may be obtained by specifying $E_\text{r} = H_\text{r} = 0$ in~\eqref{eq:NLX} and assuming normally incident and transmitted plane waves, i.e. $E = \pm\eta_0 H$, where $+$ corresponds to waves propagating in the $+z$-direction and vice-versa for $-$. To obtain the susceptibilities in~\eqref{eq:NLX}, we have to consider the transformation of two sets of independent waves. One may assume that the incident and transmitted waves for the two transformations are either both propagating in the $+z$-direction, as in Fig.~\ref{fig:NL1}, or in the $-z$-direction, as in Fig.~\ref{fig:NL2}. One may also consider the case where the waves $\Psi_1$ are propagating in the $+z$-direction and the waves $\Psi_2$ are propagating in the $-z$-direction (or vice-versa), but it may be shown that conditions for no reflection do not exist in this case.

It will next be shown that the conditions for no reflection are not the same in the two cases depicted in Figs.~\ref{fig:NL}. This is because of the presence of the square of the electric and magnetic fields in~\eqref{eq:NLX}, which introduces an asymmetry in the definition of the susceptibilities in terms of the direction of wave propagation. This asymmetry is due to the different relations between the electric and magnetic fields ($E = \pm\eta_0 H$) for forward or backward propagating waves.  Solving~\eqref{eq:NLX} for the case depicted in Fig.~\ref{fig:NL1} leads to the following conditions for no reflection:
\begin{subequations}\label{eq:RLCond}
\begin{align}
\chi^{(1)}_\text{ee} = \chi^{(1)}_\text{mm},\\
\eta_0\chi^{(2)}_\text{ee} = \chi^{(2)}_\text{mm}\label{eq:RLCond12}.
\end{align}
\end{subequations}
Similarly, the conditions for no reflection in the case depicted in Fig.~\ref{fig:NL2} read
\begin{subequations}\label{eq:RLCond2}
\begin{align}
\chi^{(1)}_\text{ee} = \chi^{(1)}_\text{mm},\\
-\eta_0\chi^{(2)}_\text{ee} = \chi^{(2)}_\text{mm}\label{eq:RLCond22}.
\end{align}
\end{subequations}
Note the minus sign difference in the relations ~\eqref{eq:RLCond12} and~\eqref{eq:RLCond22} between the second-order susceptibilities. The fact that different reflectionless metasurface conditions are obtained for different directions of propagation means that the considered nonlinear metasurface inherently exhibits a nonreciprocal response, that we will discuss in more detail later on.

As of now, we continue the evaluation of scattering from the nonlinear metasurface considering the case in Fig.~\ref{fig:NL1}. Substituting~\eqref{eq:RLCond} along with the difference of the specified fields, $\Delta E = E_\text{t} - E_\text{i}$, the average of the specified fields, $E_\text{av} = \frac{1}{2}(E_\text{t} + E_\text{i})$, and the squared average of the specified fields, $E_\text{av}^2 = E_\text{av}E_\text{av}^\ast= \frac{1}{4}(E_\text{i}^2 + E_\text{i}E_\text{t}^\ast+E_\text{t}E_\text{i}^\ast+E_\text{t}^2)$, and similarly for the magnetic fields, transforms~\eqref{eq:TDgstc2} into
\begin{equation}
\begin{split}\label{eq:ComplexEq}
2\chi^{(1)}_\text{ee} &\frac{\partial}{\partial t} E_\text{t}  + \chi^{(2)}_\text{ee} \frac{\partial}{\partial t} ( E_\text{i}E_\text{t}^\ast+E_\text{t}E_\text{i}^\ast+E_\text{t}^2) + \frac{4\eta_0}{\mu_0} E_\text{t} \\
&=\frac{4\eta_0}{\mu_0} E_\text{i} -\chi^{(2)}_\text{ee} \frac{\partial}{\partial t}E_\text{i}^2-2\chi^{(1)}_\text{ee} \frac{\partial}{\partial t} E_\text{i},
\end{split}
\end{equation}
where $E_\text{i}$ is a known excitation and $E_\text{t}$ is an unknown transmitted field. Assuming that $E_\text{i} = E_0 \cos{(\omega_0 t)}$ and that $E_0$, $\chi^{(1)}_\text{ee}$ and $\chi^{(2)}_\text{ee}$ are real quantities, corresponding to a lossless system, the relation~\eqref{eq:ComplexEq} becomes
\begin{equation}\label{eq:DiffEq}
\begin{split}
\chi^{(2)}_\text{ee} &\frac{\partial}{\partial t}E_\text{t}^2 + \left( 2\chi^{(1)}_\text{ee} +2E_0 \cos{(\omega_0 t)}\chi^{(2)}_\text{ee}\right)\frac{\partial}{\partial t} E_\text{t}  \\
&+\left(\frac{4\eta_0}{\mu_0}- 2\omega_0\chi^{(2)}_\text{ee} E_0\sin{(\omega_0 t)}\right) E_\text{t}= \frac{4\eta_0}{\mu_0} E_0 \cos{(\omega_0 t)} \\
&+\omega_0\chi^{(2)}_\text{ee} E_0\sin{(2\omega_0 t)}+2\omega_0\chi^{(1)}_\text{ee} E_0 \sin{(\omega_0 t)}.
\end{split}
\end{equation}
This is a inhomogeneous nonlinear first-order differential equation that allows one to find the transmitted field from a reflectionless birefringent second-order nonlinear metasurface with purely real susceptibilities and assuming a normally incident plane wave excitation in the $+z$-direction. The large number of assumptions that were required to obtain Eq.~\eqref{eq:DiffEq} reveals the inherent complexity of analyzing nonlinear metasurfaces.

Equation~\eqref{eq:DiffEq} does not admit an analytical solution. To obtain an approximate expression of the transmitted field, we consider that the second-order susceptibilities are much smaller than the first-order ones, which is typically a valid assumption in the absence of second order resonance, $\chi^{(2)} \approx 10^{-12} \chi^{(1)}$~\cite{boyd2003nonlinear}. From this consideration, perturbation analysis~\cite{kato2013perturbation} may be used to approximate the value of the transmitted field. Perturbation analysis stipulates that the approximate solution may be expressed in terms of a power series of the following form
\begin{equation}\label{eq:PAEt}
E_\text{t} \approx E_\text{t,0} + \epsilon E_\text{t,1} + \epsilon^2 E_\text{t,2}+...
\end{equation}
where $\epsilon$ is a small quantity. Truncating the series and solving recursively for $E_\text{t,0}$, $E_\text{t,1}$ and so on, may help reducing the complexity of the problem. Since
\begin{equation}\label{eq:PAX}
\chi_\text{ee}^{(1)} \gg \chi_\text{ee}^{(2)} \approx \epsilon,
\end{equation}
it is possible to simplify Eq.~\eqref{eq:DiffEq} using~\eqref{eq:PAEt} and solving for $E_\text{t,0}$ while neglecting all terms containing $\epsilon$. This reduces~\eqref{eq:DiffEq} to
\begin{equation}\label{eq:DiffEqEt0}
\begin{split}
2\chi^{(1)}_\text{ee}&\frac{\partial}{\partial t} E_\text{t,0}  +\frac{4\eta_0}{\mu_0}E_\text{t,0}= \frac{4\eta_0}{\mu_0} E_0 \cos{(\omega_0 t)} \\
&+2\omega_0\chi^{(1)}_\text{ee} E_0 \sin{(\omega_0 t)},
\end{split}
\end{equation}
which does not contain any nonlinear term and therefore corresponds to a simple reflectionless \emph{linear} metasurface. The steady-state solution of~\eqref{eq:DiffEqEt0} is, in complex form, given by
\begin{equation}\label{eq:Et0}
E_\text{t,0} = E_0\frac{2-jk_0\chi^{(1)}_\text{ee}}{2+jk_0\chi^{(1)}_\text{ee}}e^{j\omega_0 t},
\end{equation}
which exactly corresponds to the expected transmitted field~\cite{achouri2014general}, where the frequency of $E_\text{t,0}$ is the same as that of the incident wave. Now, $E_\text{t,1}$ can be found by inserting $E_\text{t} \approx E_\text{t,0} + \epsilon E_\text{t,1}$ into~\eqref{eq:DiffEq}, with $E_\text{t,0}$ as given in~\eqref{eq:Et0}, and neglecting all the terms containing $\epsilon^2$ (and higher powers). This leads to the following linear differential equation
\begin{equation}
\begin{split}
\chi^{(2)}_\text{ee}& \frac{\partial}{\partial t}E_\text{t,0}^2+\left( 2\chi^{(1)}_\text{ee} +2E_0 \cos{(\omega_0 t)}\chi^{(2)}_\text{ee}\right)\frac{\partial}{\partial t} E_\text{t,0} \\
&+\left(\frac{4\eta_0}{\mu_0}- 2\omega_0\chi^{(2)}_\text{ee} E_0\sin{(\omega_0 t)}\right) E_\text{t,0}+ 2\chi^{(1)}_\text{ee}\frac{\partial}{\partial t} E_\text{t,1} \\
&+ \frac{4\eta_0}{\mu_0} E_\text{t,1}  = \frac{4\eta_0}{\mu_0} E_0 \cos{(\omega_0 t)} +\omega_0\chi^{(2)}_\text{ee}E_0 \sin{(2\omega_0 t)}\\
&+2\omega_0\chi^{(1)}_\text{ee} E_0 \sin{(\omega_0 t)},
\end{split}
\end{equation}
which now contains the nonlinear susceptibility $\chi^{(2)}_\text{ee}$. This equation, of the same type as~\eqref{eq:DiffEqEt0} in $E_{\text{t},1}$, is readily solved, and yields the steady-state solution
\begin{equation}\label{eq:ET1}
\begin{split}
&E_\text{t,1}=E_0 k_0 \chi^{(2)}_\text{ee}e^{j2\omega_0 t}\\
&\cdot\left [\frac{4+12 E_0 + \chi^{(1)}_\text{ee}k_0(\chi^{(1)}_\text{ee}k_0 -4j)(E_0 -1)}{4(\chi^{(1)}_\text{ee}k_0-j)(\chi^{(1)}_\text{ee}k_0-2j)^2}\right ],
\end{split}
\end{equation}
which corresponds to second-harmonic generation, i.e. a wave at frequency $2\omega_0$, twice that of the incident wave. The procedure used to obtain $E_\text{t,1}$ may now be applied to find $E_\text{t,2}$. In this case, the differential equation becomes quite lengthy and is not shown here for the sake of conciseness. The differential equation for $E_\text{t,2}$ is also a linear first-order equation and is thus easily solved. The steady-state solution is
\begin{equation}\label{eq:Et2}
E_\text{t,2} = C_\text{t,2}\left [(2j-3\chi^{(1)}_\text{ee}k_0)e^{j\omega_0 t} + 3(2j+\chi^{(1)}_\text{ee}k_0)e^{j3\omega_0 t} \right],
\end{equation}
where the complex constant $C_\text{t,2}$ reads
\begin{equation}
\begin{split}\label{eq:Ct2}
C_\text{t,2} &=\frac{j(\chi^{(2)}_\text{ee}k_0E_0)^2}{2(\chi^{(1)}_\text{ee}k_0-2j)^3}\\
&\cdot\frac{\left [4+12 E_0 + \chi^{(1)}_\text{ee}k_0(\chi^{(1)}_\text{ee}k_0 -4j)(E_0 -1)\right ]}{(\chi^{(1)}_\text{ee}k_0-j)(\chi^{(1)}_\text{ee}k_0+2j)(3\chi^{(1)}_\text{ee}k_0-2j)}.
\end{split}
\end{equation}
The expression of $E_\text{t,2}$ corresponds to a superposition of two waves, at frequencies $\omega_0$ and $3\omega_0$. From~\eqref{eq:Ct2}, we see that the amplitude of $E_\text{t,2}$ is directly proportional to the square of $\chi^{(2)}_\text{ee}$ while the amplitude of $E_\text{t,1}$ in~\eqref{eq:ET1} is linearly proportional to $\chi^{(2)}_\text{ee}$. Similarly, $E_\text{t,2}$ is proportional to the cube of $E_0$ while $E_\text{t,1}$ is proportional to the square of $E_0$. Consequently, relations~\eqref{eq:ET1} and~\eqref{eq:Et2} remain valid only for values of $E_0$, $\chi^{(1)}_\text{ee}$ and $\chi^{(2)}_\text{ee}$ such that $E_\text{t,0} \gg E_\text{t,1} \gg E_\text{t,2}$ and $\chi^{(1)}_\text{ee} \gg \chi^{(2)}_\text{ee}$.

According to~\eqref{eq:Et0},~\eqref{eq:ET1} and~\eqref{eq:Et2}, the scattered field from the nonlinear metasurfaces may generally be expressed as
\begin{equation}\label{eq:SFNL}
E_\text{s} = \sum_{n=1}^{\infty} E_\text{s,n}e^{j n \omega_0 t},
\end{equation}
where $E_\text{s}$ represents either the reflected or the transmitted electric field and $E_\text{s,n}$ are complex constants. The form~\eqref{eq:SFNL} reveals why the synthesis of a nonlinear metasurface is not trivial: all the harmonics playing a significant role in~\eqref{eq:SFNL} should be included in the specified fields. In other words, if those harmonics were not included in the specified fields, one would not properly describe the physics of the problem. This would in fact precisely lead to the aforementioned contradiction that the susceptibilities in~\eqref{eq:NLX} would be found to be depending on time in contraction with the implicit assumption in~\eqref{eq:TDgstc} that they do not.
\begin{figure}[ht]
\centering
\subfloat[]{\label{fig:NLsimFDTD1}
\includegraphics[width=0.8\columnwidth]{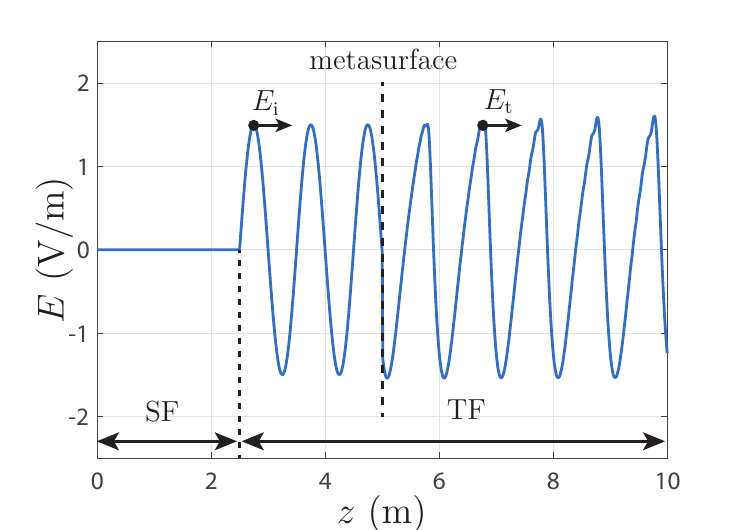}
}\\
\subfloat[]{\label{fig:NLsimFDTD3}
\includegraphics[width=0.8\columnwidth]{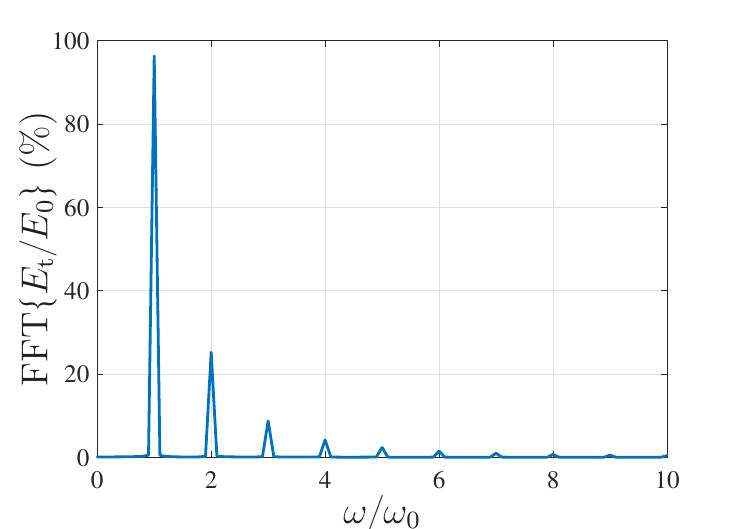}
}
\caption{FDTD simulations of a nonlinear metasurface illuminated by a plane wave from the left with the following parameters: $E_0 = 1.5$~V/m, $\chi_\text{ee}^{(1)}=0.1$~m and $\chi_\text{ee}^{(2)}=0.004$~m$^2$/V. (a)~Electric field waveform. (b)~Spectrum.}
\label{fig:NLsimFDTD}
\end{figure}

We shall now validate our theory with full-wave analysis. The metasurface is synthesized so as to satisfy the conditions for no reflection given in~\eqref{eq:RLCond}, and we consider the following arbitrary parameters: $E_0 = 1.5$~V/m, $\chi_\text{ee}^{(1)}=0.1$~m and $\chi_\text{ee}^{(2)}=0.004$~m$^2$/V. The simulations are performed with the 1D FDTD zero-thickness metasurface simulation code developed in~\cite{vahabzadeh2017space},  modified to account for the nonlinear susceptibilities (see Appendix~\ref{sec:NLFDTD})~\footnote{Normalized constants are used in all the simulations, so that $c_0 = \epsilon_0 = \mu_0 = f_0 = 1$.}.

In the first simulation, whose results are plotted in Fig.~\ref{fig:NLsimFDTD1}, the metasurface is illuminated by a plane wave propagating in the $+z$-direction. In the figure, the metasurface is placed at the center and the simulation area is split into a scattered-field region~(SF) and a total-field region~(TF). The source is placed at the SF/TF boundary on the left of the metasurface. As expected, the metasurface is reflectionless ($E=0$ in the SF region). A time-domain Fourier transform of the steady-state transmitted field is performed and the normalized (to $E_0$) result is plotted in Fig.~\ref{fig:NLsimFDTD3}. It may be seen that the metasurface generates a transmitted field with several visible harmonics satisfying $E(\omega_{n})>E(\omega_{n+1})$, as expected.

\begin{figure}[ht]
\centering
\subfloat[]{\label{fig:NLsimFDTD2}
\includegraphics[width=0.8\columnwidth]{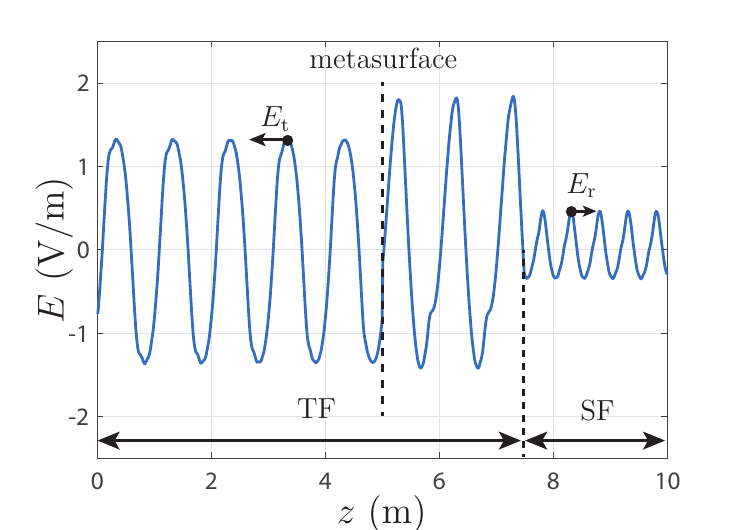}
}\\
\subfloat[]{\label{fig:NLsimFDTD4}
\includegraphics[width=0.8\columnwidth]{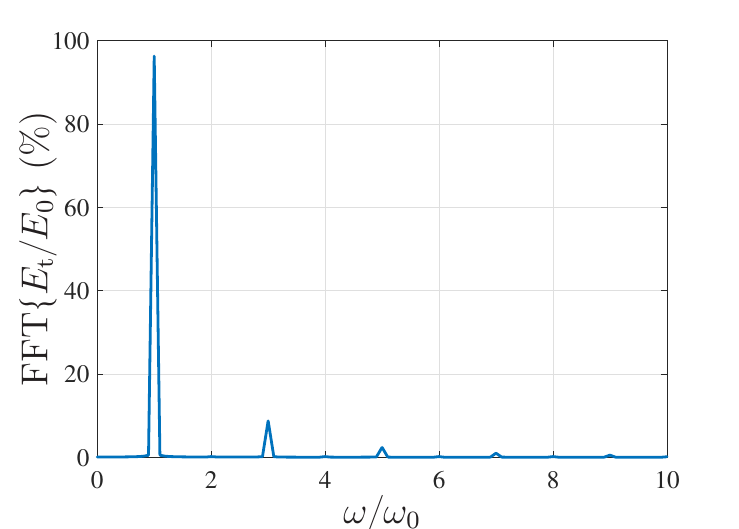}
}
\caption{FDTD simulations of a nonlinear metasurface illuminated by a plane wave from the right with the following parameters: $E_0 = 1.5$~V/m, $\chi_\text{ee}^{(1)}=0.1$~m and $\chi_\text{ee}^{(2)}=0.004$~m$^2$/V. (a)~Electric field waveform. (b)~Spectrum.}
\label{fig:NLsimFDTD22}
\end{figure}

In order to investigate its nonreciprocity, the metasurface is now illuminated from the left by a plane wave propagating in the $-z$-direction. The corresponding simulated waveform is plotted in Fig.~\ref{fig:NLsimFDTD2}, where the positions of the SF/TF regions have been changed accordingly. We can see that when this metasurface is illuminated from the right, it is not reflectionless anymore as evidenced by the nonzero electric field in the SF region. This is in agreement with the fact that different conditions for no reflection apply for different directions of propagation, according to the discussion that led to Eqs.~\eqref{eq:RLCond} and~\eqref{eq:RLCond2}.

The time-domain Fourier transform of the transmitted field in Fig.~\ref{fig:NLsimFDTD2} is plotted in Fig.~\ref{fig:NLsimFDTD4}. As may be seen, the transmitted field is missing frequencies that are even multiples of $\omega_0$. In fact, these missing frequencies are reflected, since the system is assumed to be lossless, by the metasurface instead of being transmitted. Thus, the nonreciprocal behavior of the second-order nonlinear metasurface only affects frequencies that are even multiples of $\omega_0$.
\begin{figure}[ht]
\centering
\includegraphics[width=0.8\columnwidth]{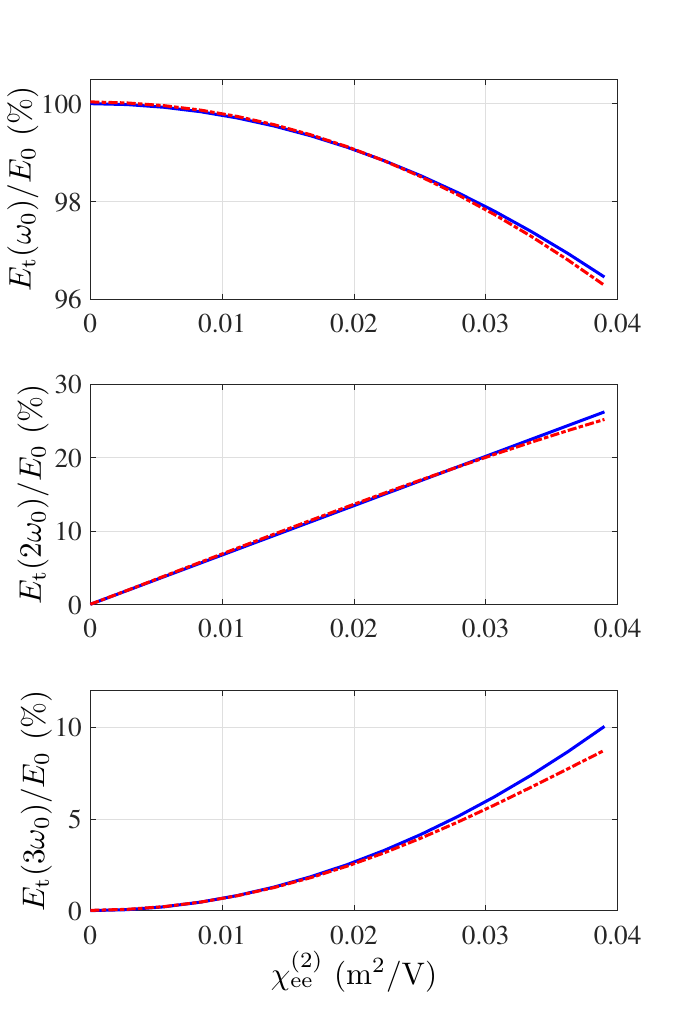}
\caption{Comparisons of the amplitudes of the first three normalized harmonics obtained by theory [Eqs.~\eqref{eq:Et0},~\eqref{eq:ET1} and~\eqref{eq:Et2}] (blue solid lines) and FDTD simulation (red dashed lines) for the following parameters:~$E_0 = 1.5$~V/m, $\chi_\text{ee}^{(1)}=0.1$~m.}
\label{fig:NLsim1}
\end{figure}

Next, we shall compare the theoretical results with FDTD simulations. For this purpose, two metasurfaces  satisfying~\eqref{eq:RLCond} with different susceptibilities are considered. The values of $E_0$ and $\chi_\text{ee}^{(1)}$ are specified while those of $\chi_\text{ee}^{(2)}$ are swept. The amplitudes of the first three harmonics (at $\omega_0$, $2\omega_0$ and $3\omega_0$) of the transmitted field are obtained from~\eqref{eq:Et0},~\eqref{eq:ET1} and~\eqref{eq:Et2}, and compared to the corresponding amplitudes found by FDTD simulation. Note that the amplitude of the harmonic at $\omega_0$ is computed from both~\eqref{eq:Et0} and~\eqref{eq:Et2} since both of these equations include terms contributing to this harmonic. As explained in the Appendix~\ref{sec:NLFDTD}, the FDTD field update equations are only valid for specific values of $E_0$, $\chi_\text{ee}^{(1)}$ and $\chi_\text{ee}^{(2)}$, other specifications leading to a nonphysical behavior. In addition to choosing those parameters so as to follow this constraint, we have ensured $\chi_\text{ee}^{(1)} > \chi_\text{ee}^{(2)}$ in order to be consistent with the assumptions of the perturbation analysis method [see Eq.~\eqref{eq:PAX}].

\begin{figure}[ht]
\centering
\includegraphics[width=0.8\columnwidth]{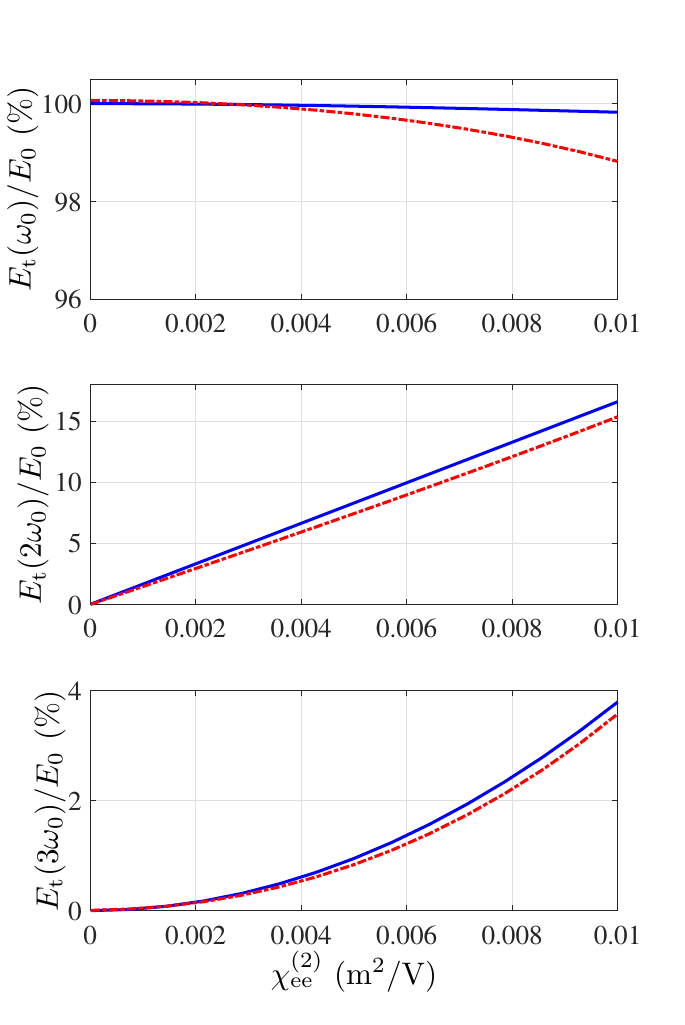}
\caption{Comparisons of the amplitudes of the first three normalized harmonics obtained by theory [Eqs.~\eqref{eq:Et0},~\eqref{eq:ET1} and~\eqref{eq:Et2}] (blue solid lines) and FDTD simulation (red dashed lines) for the following parameters:~$E_0 = 10$~V/m, $\chi_\text{ee}^{(1)}=0.3$~m.}
\label{fig:NLsim2}
\end{figure}

The first comparison, presented in Fig.~\ref{fig:NLsim1}, considers $E_0 = 1.5$~V/m, $\chi_\text{ee}^{(1)}=0.1$~m and $\chi_\text{ee}^{(2)}$ swept in $[0,0.04]$~m$^2$/V, while the second comparison, presented in Fig.~\ref{fig:NLsim2}, considers $E_0 = 10$~V/m, $\chi_\text{ee}^{(1)}=0.3$~m and $\chi_\text{ee}^{(2)}$ swept in $[0,0.01]$~m$^2$/V. Both comparisons show good agreement between theory and simulation. In both cases, the discrepancies between the results of the two methods increase with increasing $\chi_\text{ee}^{(2)}$, as expected from the perturbation assumption $\chi_\text{ee}^{(1)} > \chi_\text{ee}^{(2)}$. Thus, as $\chi_\text{ee}^{(2)}$ increases towards $\chi_\text{ee}^{(1)}$, the error in the approximation of the transmitted field (Eq.~\eqref{eq:PAEt}) progressively increases. Another source of error is the fact that Eq.~\eqref{eq:PAEt} is truncated at the third term and that higher-order terms are neglected, which induces additional errors.

We have verified that the FDTD simulations in Figs.~\ref{fig:NLsimFDTD} and~\ref{fig:NLsimFDTD22}, as well as in Figs.~\ref{fig:NLsim1} and~\ref{fig:NLsim2} satisfy the losslessness and passivity power conservation condition \mbox{$\sum_{n=1}^\infty |E(\omega_n)|^2 = |E_0|^2$}. In contrast, this condition does not hold in the case of the theoretical results, due to the truncation of~\eqref{eq:PAEt}, as is clearly apparent in Fig.~\ref{fig:NLsim2}, where $\sum_{n=1}^3|E(\omega_n)|^2 > |E_0|^2$ for values of $\chi_\text{ee}^{(2)}$ that are close to 0.01~m$^2$/V.

\section{Conclusion}

We have investigated a particular case of second-order nonlinear isotropic metasurfaces that possess both electric and magnetic nonlinear susceptibilities. We have found the synthesis expressions relating the susceptibilities to the fields on both sides of the metasurface, which lead to the derivation of the reflectionless metasurface conditions. These conditions reveal the inherent nonreciprocal nature of nonlinear metasurfaces. Then, the scattered field from such metasurfaces was analyzed based on perturbation theory as well as full-wave simulations, and good agreement was found between the two approaches.

\section*{Acknowledgment}

This work was accomplished in the framework of the Collaborative Research and Development Project CRDPJ 478303-14 of the Natural Sciences and Engineering Research Council of Canada (NSERC) in partnership with the company Metamaterial Technology Inc.

\appendix
\section{Finite-Difference Time-Domain Scheme for Nonlinear Metasurfaces}
\label{sec:NLFDTD}

Here, we extend the 1D finite-difference time-domain (FDTD) simulation scheme developed in~\cite{vahabzadeh2017space} for the analysis of metasurfaces to the case of nonlinear susceptibility components. This FDTD scheme consists in using traditional FDTD update equations everywhere on the simulation grid except at the nodes positioned before and after the metasurface. For these specific nodes, the update equations are modified, using the GSTCs relations, to take into account the effect of the metasurface. The conventional Yee-grid FDTD 1D equations are given by
\begin{subequations}\label{eq:UDE}
\begin{equation}
H_y^{n+1}(i) = H_y^{n}(i) - \frac{\Delta t}{\mu_0\Delta z}\left(E_x^{n+\frac{1}{2}}(i+1)  - E_x^{n+\frac{1}{2}}(i)  \right),
\end{equation}
\begin{equation}
E_x^{n+1/2}(i) = E_x^{n-\frac{1}{2}}(i) - \frac{\Delta t}{\epsilon_0\Delta z}\left(H_y^{n}(i)  - H_y^{n}(i-1) \right),
\end{equation}
\end{subequations}
where $i$ and $n$ correspond to the cell number and time coordinates and $\Delta z$ and $\Delta t$ are their respective position and time steps. The metasurface is placed at a virtual position between cell number $i=n_d$ and $i=n_d+1$, corresponding to a position between an electric node and a magnetic node. To take into account its effect, a virtual electric node is created just before the metasurface (at $i=0^-$) and a virtual magnetic node is created just after the metasurface (at $i=0^+$).

From~\eqref{eq:UDE}, the update equations for $H_y^{n+1}(n_d)$ and $E_x^{n+\frac{1}{2}}(n_d+1)$ are connected to these virtual nodes via the following relations
\begin{subequations}\label{eq:UDE2}
\begin{equation}
H_y^{n+1}(n_d) = H_y^{n}(n_d) + \frac{\Delta t}{\mu_0\Delta z}\left(E_x^{n+\frac{1}{2}}(0^-)  - E_x^{n+\frac{1}{2}}(n_d)  \right),
\end{equation}
\begin{equation}
\begin{split}
E_x^{n+\frac{1}{2}}(n_d+1) &= E_x^{n-\frac{1}{2}}(n_d+1) \\
&+ \frac{\Delta t}{\epsilon_0\Delta z}\left(H_y^{n}(n_d+1)  - H_y^{n}(0^+)  \right),
\end{split}
\end{equation}
\end{subequations}
where the value of the electric and magnetic fields at the virtual nodes are obtained from the GSTCs relations
\begin{subequations}\label{eq:GSCTsY}
\begin{align}
-\Delta H_y &= \epsilon_0 \chi_\text{ee}^{(1)}\frac{\partial}{\partial t} E_{x,\text{av}} + \epsilon_0 \chi_\text{ee}^{(2)}\frac{\partial}{\partial t} E_{x,\text{av}}^2,\\
-\Delta E_x &= \mu_0 \chi_\text{mm}^{(1)}\frac{\partial}{\partial t} H_{y,\text{av}} + \mu_0 \chi_\text{mm}^{(2)}\frac{\partial}{\partial t} H_{y,\text{av}}^2.
\end{align}
\end{subequations}
Using~\eqref{eq:GSCTsY}, the expression of the electric and magnetic fields at the virtual nodes in~\eqref{eq:UDE2} read
\begin{subequations}\label{eq:fdtdGSTC}
\begin{equation}
\begin{split}
&H_y^{n}(0^+) = H_y^{n}(n_d) - \frac{\epsilon_0\chi_\text{ee}^{(1)}}{\Delta t}\left(E_{x,\text{av}}^{n+\frac{1}{2}}  - E_{x,\text{av}}^{n-\frac{1}{2}} \right)\\
&- \frac{\epsilon_0\chi_\text{ee}^{(2)}}{\Delta t}\left((E_{x,\text{av}}^{n+\frac{1}{2}})^2  - (E_{x,\text{av}}^{n-\frac{1}{2}})^2 \right),
\end{split}
\end{equation}
\begin{equation}
\begin{split}
&E_x^{n+\frac{1}{2}}(0^-) = E_x^{n+\frac{1}{2}}(n_d+1) + \frac{\mu_0\chi_\text{mm}^{(1)}}{\Delta t}\left(H_{y,\text{av}}^{n+1}  - H_{y,\text{av}}^{n}  \right) \\
&+ \frac{\mu_0\chi_\text{mm}^{(2)}}{\Delta t}\left((H_{y,\text{av}}^{n+1})^2  - (H_{y,\text{av}}^{n})^2  \right),
\end{split}
\end{equation}
\end{subequations}
where the average electric field is defined by
\begin{equation}\label{eq:avfdtd}
E_{x,\text{av}}^{n+\frac{1}{2}} = \frac{E_x^{n+\frac{1}{2}}(n_d) + E_x^{n+\frac{1}{2}}(n_d+1)}{2},
\end{equation}
and similarly for the average magnetic field. Substituting~\eqref{eq:fdtdGSTC} along with~\eqref{eq:avfdtd} into~\eqref{eq:UDE2} leads to two quadratic equations that may be independently solved to obtain the final update equations. These two quadratic equations yield, each one of them, two possible solutions but only one of the two correspond to a physical behavior. The two solutions that produce physical results are
\begin{subequations}\label{eq:FinalUDE}
\begin{equation}
H_y^{n+1}(n_d) = \frac{(2\Delta z -\chi_\text{ee}^{(1)} - H_y^{n+1}(n_d+1)\chi_\text{ee}^{(2)})\mu_0 - \sqrt{\Delta_\text{h}}}{\mu_0\chi_\text{ee}^{(1)}},
\end{equation}
\begin{equation}
E_x^{n+\frac{1}{2}}(n_d+1) = \frac{(2\Delta z -\chi_\text{mm}^{(1)} - E_x^{n+\frac{1}{2}}(n_d)\chi_\text{mm}^{(2)})\epsilon_0 - \sqrt{\Delta_\text{e}}}{\epsilon_0\chi_\text{mm}^{(1)}}
\end{equation}
\end{subequations}
where the discriminant $\Delta_\text{h}$ is given by
\begin{equation}
\begin{split}
\Delta_\text{h} =& \mu_0\Big\{4 \Delta t(E_x^{n+\frac{1}{2}}(n_d)-E_x^{n+\frac{1}{2}}(n_d +1))\chi_\text{ee}^{(2)}+ \mu_0\Big[4 \Delta z^2 \\
&+ (\chi_\text{ee}^{(1)}+(H_y^{n}(n_d)+H_y^{n+1}(n_d+1))\chi_\text{ee}^{(2)})^2 \\
&-4 \Delta z (\chi_\text{ee}^{(1)} +(H_y^{n}(n_d)+H_y^{n+1}(n_d+1))\chi_\text{ee}^{(2)} )  \Big]\Big\},
\end{split}
\end{equation}
and the discriminant $\Delta_\text{e}$ is given by
\begin{equation}
\begin{split}
\Delta_\text{e} =& \epsilon_0\Big\{4 \Delta t(H_y^{n}(n_d)-H_y^{n}(n_d +1))\chi_\text{mm}^{(2)}+ \epsilon_0\Big[4 \Delta z^2 \\
&+ (\chi_\text{mm}^{(1)}+(E_x^{n-\frac{1}{2}}(n_d)+E_x^{n-\frac{1}{2}}(n_d+1))\chi_\text{mm}^{(2)})^2 \\
&-4 \Delta z (\chi_\text{mm}^{(1)} +(E_x^{n+\frac{1}{2}}(n_d)+E_x^{n-\frac{1}{2}}(n_d+1))\chi_\text{mm}^{(2)} )  \Big]\Big\}.
\end{split}
\end{equation}

Because of the square roots in~\eqref{eq:FinalUDE}, the update equations may lead to nonphysical behavior depending on the values of the two discriminants. This limits the range of allowable values that the susceptibilities and the amplitude of the incident field may take.

\bibliography{LIB}

\end{document}